\documentclass[10pt,conference]{IEEEtran}
\IEEEoverridecommandlockouts
\usepackage{cite}
\usepackage{amsmath,amssymb,amsfonts}
\usepackage{algorithmic}
\usepackage{graphicx}
\usepackage{textcomp}
\usepackage{xcolor}
\def\BibTeX{{\rm B\kern-.05em{\sc i\kern-.025em b}\kern-.08em
    T\kern-.1667em\lower.7ex\hbox{E}\kern-.125emX}}

\begin{document}

\title{Software Logging for Machine Learning}

\author{\IEEEauthorblockN{Nathan Bosch}
\IEEEauthorblockA{\textit{Development Unit Radio Services \& Infra Analytics (DSI Analytics)} \\
\textit{Ericsson AB}\\
Gothenburg, Sweden \\
nathan.bosch@ericsson.com}
\and
\IEEEauthorblockN{Jan Bosch}
\IEEEauthorblockA{\textit{Department of Computer Science and Engineering} \\
\textit{Chalmers University of Technology}\\
Gothenburg, Sweden \\
jan@janbosch.com}

}

\maketitle

\begin{abstract}
System logs perform a critical function in software-intensive systems as logs record the state of the system and significant events in the system at important points in time. Unfortunately, log entries are typically created in an ad-hoc, unstructured and uncoordinated fashion, limiting their usefulness for analytics and machine learning. In this paper, we present the main challenges of contemporary approaches to generating and storing system logs data for large, complex, software-intensive systems based on an in-depth case study at a world-leading telecommunications company. Second, we present a systematic and structured approach for generating log data that does not suffer from the aforementioned challenges and is optimized for use in machine learning. Third, we provide validation of the approach based on expert interviews that confirms that the approach addresses the identified challenges and problems.
\end{abstract}


\section{Introduction}

Logging is a critical activity in software-intensive systems as it records the state of the system and significant events in the system at important points in time. This facilitates significantly simplified defect management, anomaly detection, monitoring of system performance over time and even prediction of future system behavior. As a consequence, the use of system logs is ubiquitous for software intensive systems and the exchange of log files between users of these systems and the organizations that build them is a typical activity, especially when there is (circumstantial) evidence of the system not performing as expected.

Despite their ubiquitous and highly informative nature, the full potential of system logs is not realized in most organizations. The primary reason is that most logs are intended for human interpretation but, because of constantly increasing system size and complexity, have grown to an unmanageable size. It is not atypical for log files to measure in the gigabytes of data for even a single day of logging and it is virtually impossible to analyse the data solely with human means.

The logical response to the aforementioned challenge is to employ machine learning techniques, but the unstructured, alphanumerical and human oriented nature of logs limits the applicability of machine learning algorithms, as there typically is quite some noise and irrelevant information in system logs. This causes a need for significant pre-processing of logs which requires large amounts of human effort, reducing the benefits that fully automated machine learning could provide.

The most promising way to address the aforementioned challenge is to focus on the way system logs and the entries in those logs are generated. Doing so in the right way could remove the need for pre-processing and opens up direct, potentially even automated, use of system logs for training of machine learning models and inference based on the information in logs.

The contribution of this paper is threefold. First, we present the main challenges of contemporary approaches to generating and storing log data for large, complex, software-intensive systems based on an in-depth case study at a world-leading telecommunications company. Second, we present a systematic and structured approach for generating log data that does not suffer from the aforementioned challenges and that allows for immediate use by machine learning algorithms. Third, we provide validation based on expert interviews that confirms that the approach addresses the identified challenges and problems.

The remainder of this paper is organized as follows. In the next section, we present the research method that we used in this study. Section~\ref{background} presents the background and related work. This is followed by section~\ref{problem} presents the problem statement. After which, in section~\ref{solution}, we present the application scenarios, technical approach and process approach. We provide validation of our approach in section~\ref{validation}. Finally, we provide a discussion and a conclusion in sections~\ref{discussion} and~\ref{conclusion}, respectively.

\section{Research method}
The research reported in this paper is based on design science~\cite{hevner07} as an overall research method. Design science focuses on three cycles, i.e. the relevance cycle, the design cycle and the rigor cycle. The relevance cycle in our research focused on studying the challenges associated with using systems logs of different types for machine learning, although the primary focus was on anomaly detection.

The research was conducted at a primary case study company through a participant-observer approach as well as secondary cases from other companies through action research. As part of the work at the primary case company, we studied a variety of different types of logs. Specifically, we studied logs for two generations of systems. For each generation, we studied two major subsystems and, for each subsystem, we studied both operational and error logs. All types of logs were generated in two contexts, i.e. during internal testing and at customer sites. The log data investigated as part of the research covered more than 12 months of log data collection. Although the primary form of data collection during the research was through observation, the first author also conducted interviews with domain experts. The results of the relevance cycle are presented in section~\ref{problem}.

The rigor cycle is concerned with relating to the knowledge base. We present the background and related work in section~\ref{background}. In section~\ref{discussion}, we present our scientific contribution as an addition to the knowledge base.

The design cycle is concerned with developing a solution for the identified challenges. The proposed approach to generating logs for machine learning is presented in section~\ref{solution}. Our approach consists of the definition of a set of scenarios for system log usage as well as a technical solution for generating logs and a process defining how R\&D teams should work with logs. 

We validated the proposed approach using expert feedback. First, we conducted a validation workshop with a group of eight experts at the primary case study company where we presented the identified challenges as well as the technical solution and process approach presented in this paper. Second, we conducted follow up interviews with some of the experts after the workshop based on their interest and explicit request.

The research at the primary case company was conducted between June 2019 and January 2020. During the period June 2019 - August 2019 the first author was working as an employee at the company full time and during the period September 2019 - December 2019 the first author worked part time. The second author has studied, among others, software logging at a variety of companies over the years, among others in the context of Software Center~\footnote{www.software-center.se}, and is involved with the primary case company through a variety of research projects and has collaborated with the organization since 2011.

The primary case study company is a world-leading provider of telecommunications solutions that offers a variety of products, solutions and services. For the primary case as well as the secondary cases, we are, for reasons of confidentiality, unable to provide details on the specific systems that we studied nor on the customers involved in the research.

\section{Background \& Related work} \label{background}



As systems and their complexity grow, the need for automatic evaluation of the system during operation grows with it. Large software intensive systems companies use system logs as one of the primary methods for evaluating system performance. There is existing research concerning the analysis of system logs, specifically concerning detection of anomalies in system behaviour and performance. The types of models are extensive, among the more common ones: workflow methods, such as~\cite{Lou2010} and~\cite{Yu2016}, rule-based methods, such as~\cite{Cinque2013}, PCA methods, such as~\cite{Xu2008} and~\cite{Xu2009}, and - more recently - LSTM language modelling, such as~\cite{Du2017} and~\cite{Brown2018}.

While the basic approach of each method of log analysis and anomaly detection is different, each approach is, to a varying degree, reliant on a structured decomposition of log entries. While the results are often quite successful, rule-based methods require extensive domain knowledge and significant effort to build effective systems. Methods that apply unsupervised machine learning techniques, such as~\cite{Xu2009}, to achieve their results are less dependent on expert domain knowledge, but still require either source code analysis or significant amounts of parsing for useful feature extraction. Our research shows that this frequently is not feasible as source code is often not available for analysis and as many logs are highly unstructured and consequently difficult to parse.

The problems concerning access to source code~\cite{Messaoudi2018} and lack of structure~\cite{Du2017} have been acknowledged in research and, due to the frequency of this problem, automated log parsers, such as MoLFI~\cite{Messaoudi2018}, Drain~\cite{He2017}, and Spell~\cite{Du2017Spell}, have been developed to some success. These parsers are built to infer the true, underlying log events for each log entry as well as extract parameter information. An analysis of 13 different automated log parsers was done by Zhu et al.~\cite{Zhu2019} where the analysed automated log parsers were organized into 4 different categories: frequent pattern mining, clustering, heuristics and other approaches. Each of the 13 automated log parsers was tested on a series of datasets, selected from their loghub data repository. Despite the standardisation of the logs, and consequently the relatively low variance in structure of the data on which the performance of the automated log parsers were tested, each of the automated log parsers struggled to reach 80\% accuracy at correctly parsing log messages for at least six of the 16 datasets. Even the state-of-the-art log parser, Drain~\cite{He2017}, struggles with state identification and dealing with log messages of variable length~\cite{Zhu2019}. Although impressive as research results, in practice many companies have logs with less standardization and our research shows that automated log parsing does not provide the desired results.





\section{Problem Statement}\label{problem}
Although logging may seem a trivial activity, in the case of large, complex production systems that have high reliability requirements, the creation, analysis and interpretation of system logs easily becomes a complicated activity. This is because there are several or dozens of teams involved that are all trying to track the functionality that each team is working on. This leads to a high degree of variance in the way that logs are generated. Second, there are frequently multiple log files that teams can use for generating log entries of different types. When logs are interpreted by developers, this can serve as a strong solution to the first point, as the separation of functionality is much more useful for determining faults by developers. In a machine learning context, however, this separation can greatly complicate the process of processing the necessary data. Finally, the logs and log entries evolve in an unmanaged and unpredictable fashion. Although the contemporary approach to generating system logs is fine for human interpretation of logs, as humans are very versatile in interpreting information, when using machine learning for analysing and interpreting logs, the consequence is that these logs prove difficult to analyse and that automatic interpretations easily become incorrect.


The challenge of applying machine learning to system logs is exacerbated in the case where the system is subject to continuous deployment of new software versions. In that case, teams in R\&D even more frequently change the way logging is performed which adds an evolution challenge to the already complicated activity of logging.

\begin{figure}\centering
\includegraphics[width=9cm, keepaspectratio]{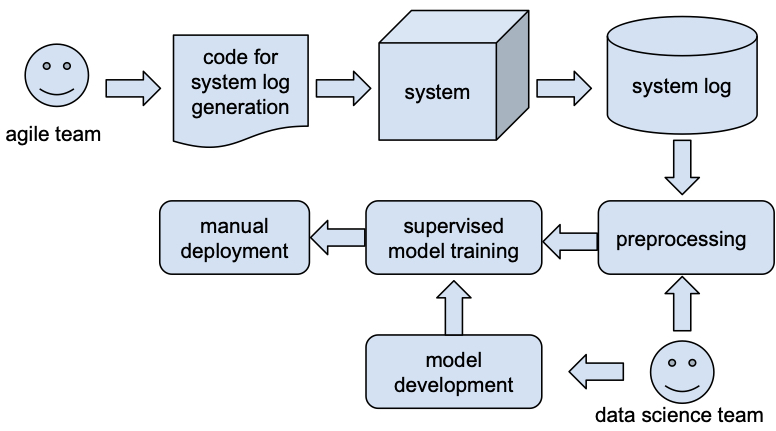}
\caption{Traditional system logging approach}
\label{TradSysLog}
\end{figure}

The traditional way of generating system logs is shown in figure~\ref{TradSysLog}. In this case, the R\&D teams have full freedom to generate logs to optimally support their needs. These approaches often have developed over time to optimally support developers. The challenge is that when the same logs are used for machine learning, the data science team is required to spend significant effort on pre-processing, the pre-processed data then used to manually (re)train the model which is, subsequently, manually deployed.

Based on the literature that we have studied, to the best of our knowledge there are no companies that have fully addressed the challenge of applying machine learning to system logs as this is a relatively new use case, caused by the recent emergence of ML/DL. 

As part of our case study research at the primary and secondary case companies as well as based on the literature that we reviewed and reported in section~\ref{background}, we have identified eight significant challenges associated with applying machine learning to system logs. Below we discuss each challenge in more detail.

\noindent(i) {\bf Logs require extensive pre-processing}: The first challenge is that there are few, if any, constraints on or guidelines for how developers should generate log entries, mainly because there has not been a need to do so until recently. This leads to log files that are highly diverse in content and semantics. The result is that any analysis, interpretation or inference based on log files requires extensive pre-processing before any automation can be applied. This problem is derived from the literature. Several researchers have reported on the difficulty using and writing regex parsers (\cite{Messaoudi2018},~\cite{He2017},~\cite{Du2017Spell}) and proposed the use of automated log parsers. Automated log parsing is, however, brittle to changes in structure and varying lengths of message~\cite{Zhu2019}.

\noindent(ii) {\bf Multiple processes write into the same log file}: Especially for machine learning, one critical source of information is the sequence in which data is presented to the algorithm, i.e. time series. However, in the cases that we studied it was quite common for multiple processes to write into the same log file. This causes a situation where it becomes significantly more complicated to treat logs as time series (or at least this adds significant noise).  Using identifiers and other solutions would provide a way around this, but that requires a coordination between R\&D teams that often does not exist. This is no problem in the case of traditional log analysis by humans but it does complicate using system logs for machine learning.

\noindent(iii) {\bf Comprehensive information about the system spread over multiple log files}: Development teams frequently decide to use different log files for different types of logging as it helps them to interpret the data from each log. In the case of machine learning, however, spreading relevant system information over multiple system log files significantly complicates the training process. This is because it requires combining pre-processed and analysed data from multiple logs in order to reach reliable conclusions and to infer useful classifications or predictions.

\noindent(iv) {\bf Different log files contain information at fundamentally different abstraction levels}: In multiple cases, we identified that, especially for large systems, R\&D teams frequently generate different types of logs for data at different abstraction levels. This can express itself in multiple dimensions, including time (e.g. milliseconds, seconds, minutes, hours, days or weeks), aggregation level (e.g. component, subsystem or system level) and instance versus population viewpoints. Although this is generally the most effective approach for human interpretation, in the case of machine learning this provides a challenge. Essentially, the difference in abstraction level of the information in these log files complicates automated training and inference based on log data.

\noindent(v) {\bf Interpreting log file data requires extensive domain knowledge}: To build a well-performing machine learning model, the data scientist or engineer needs to understand the domain well. This is required to be able to select the correct features, to evaluate data sets to understand correlations between data items, etc. The challenge is that those working with machine learning tend to lack the domain knowledge and log files already represent a significant abstraction and to some extent obfuscation of the actual system. Conversely, the engineers that have significant domain expertise tend to be limited in number, highly sought after and often lack a good understanding of machine learning.

This problem has been presented in several articles. Some machine-learning techniques for log analysis/parsing take this lack of domain knowledge into account (e.g.~\cite{Du2017},~\cite{Du2017Spell}), or at least recognise that it presents challenges~\cite{yuan2012characterizing}. However, with a hard to interpret log, the output will also be difficult to interpret. 

\noindent(vi) {\bf Developers add code that generates log data in ad-hoc formats}: One of the key challenges that is a root cause for the aforementioned challenges is that entries in system log files are generated with no or minimal constraints on the content. This is a desirable and attractive element of system log generation in the case of interpretation by developers. For machine learning purposes, the notion of a log entry that can range from a few bytes to many megabytes is complicated to handle. The enormous variety in the structure and size of data even in the same log file significantly complicates automated interpretation and inference using machine learning.

\noindent(vii) {\bf Unmanaged evolution of log data generation invalidates cross-log file training}: In a DevOps environment, developers typically are free to change the way log data is generated on a continuous basis. This results in a situation where older log files cannot be used for training in the context of machine learning without compensating for the change in log data generation.

\noindent(viii) {\bf Automated log file interpreters fail unpredictably due to changed log data structure}: Finally, as a consequence of the aforementioned challenges, automated parsers used for pre-processing log files tend to stop working or to generate incorrect output because of the unpredictable evolution of the structure of log files. As mentioned in the background, the automated log parsers do not perform very well on complicated logs~\cite{Zhu2019}. These automated log parsers require consistent structure for the same log entry, otherwise it will recognise the newly structured entry as a new entry entirely.

Reflecting on the challenges in this section, our main observation is that system logs have been developed for interpretation by engineers and R\&D teams. The structure of these logs is suitable for these purposes. However, if one desires to use system logs for machine learning, a different approach must be followed. There are at least four main implications that need to be addressed by any solution approach. First, the structure of log files and log entries need to follow a defined framework and R\&D teams need to adhere to this. Second, the log file should be used for the intended purpose of recording the state of the system and significant events. Third, the data should be generated such that the need for pre-processing of data is minimized or, preferably, entirely removed. Finally, there is a need to instantiate software processes to support logging based on these principles to allow for the use of machine learning for classifying and predicting system behaviour and detecting anomalies. Despite extensive exploration, we have found no research on guidelines or practices for generating logs which would serve these purposes. For a company developing and deploying mission critical, complex software intensive systems, this presents a major bottleneck.

\section{System logs for machine learning}\label{solution}

So far in this paper, we have presented the limitations of using system logs intended for use by engineers for machine learning. In addition, we have presented the background and related work and concluded that the current state of the art does not provide a viable approach for system logs optimized for machine learning. In this section, we present our approach that addresses this concern. Our approach consists of three main parts. First, we discuss the typical scenarios that logs optimized for machine learning could be applied to and the success factors which would emerge from each scenario.  Second, we propose a more detailed technical implementation. Finally, assuming a traditional or DevOps R\&D context, we present the changes to software processes required for realizing the proposed approach in an industrial context in order to provide system logs specific for machine learning.

\subsection{Scenarios for Machine Learning Logs}
Based on our research at the case study company, as well as experience from other companies, we identified that there are at least three different scenarios in which system logs for machine learning are used, i.e. the traditional scenario, the DevOps scenario and the autonomous system scenario.

First, the traditional case is where a data scientist receives a system log for analysis of specific aspects or factors. This is the main scenario in cases where unique challenges or opportunities are identified. In this case, a data science expert is brought in to focus on a particular question or topic. The data scientist then selects the relevant information from the system log, i.e. feature engineering, as well as, where relevant, collects information from other sources with the intent of providing quantitative insight into the data.

The second scenario is where system logs are used in a DevOps environment. In this case, agile teams receive the system logs from the latest deployment to determine the impact of their development efforts, identify issues or the resolution of issues, track the impact on system properties, etc. The advantage of receiving system logs frequently is that it allows teams and the organization as a whole to quantitatively establish the value of the development efforts by the R\&D organization. For example, in earlier research~\cite{bosch17} we have shown that half or more of all new features added to a system are hardly, if ever, used. For features developed over multiple sprints, teams can use this information to decide whether to continue to develop the feature, halt development or even remove the feature.

The third scenario is the fully autonomous case, which is concerned with system deployments where the system autonomously monitors its own behaviour and retrains its machine learning models when required. The monitoring of system behaviour could trigger warnings and retrain the models when they start to perform poorly or when the system acts outside predefined boundaries. Although this may seem far fetched, our research shows that companies that employ experimentation techniques such as A/B testing and multi-armed bandits are increasingly willing to let systems experiment autonomously with their behaviour in order to continuously improve their performance.

There generally are three factors that are relevant, i.e. effort, data relevance and automation. Effort is concerned with the amount of time that engineers and data scientists need to spend on system logs in order to derive relevant and actionable data from the system. Data relevance refers to the timeliness of the data in system logs. The older the data in the logs, the more time has passed. In the mean time, the system, as well as its context, may have changed, which increases the likelihood of deriving incorrect conclusions due to outdated data. Finally, ideally we would like to remove humans from the entire process, which requires increased automation. Automation requires a high degree of standardization in the way system logs are generated and the timing of log delivery.

For {\bf traditional logs} the primary success factor is to reduce the manual effort required to derive insights from system logs. The approach presented in this paper delivers on this as we generate logs in such a way that there is no need to pre-process the data. This requires collaboration between engineers and data scientists during the development of the code for generating system log entries. A second advantage is that agreeing on a unified log representation reduces effort for data science teams as the pre-processing does not have to be repeated multiple times.

In the case of a {\bf DevOps environment} all key success factors are relevant, i.e. lower effort, data relevance and automation. The arguments concerning effort are the same as for traditional logs. Data relevance is important in this case as the software in the system is deployed frequently. As a consequence, the system logs need to be carefully selected to make sure that the data in the log is still relevant and representative for the current system. Typically, the DevOps teams are most concerned with the logs from the last deployment, but for specific types of analysis teams may use a time window over system logs from the most recent software deployments. As data pipelines now need to be explicitly managed to make them more robust to changes, the generation of system logs should be centralised for the system as this allows for easier resolution of issues. For instance, significant changes in parameter information will be easily detected which then allows for standardised logs that don't break the machine learning deployments. As the analysis is conducted for every deployment, there is a significant need to automate the analysis of data as it would otherwise easily become overly effort consuming for DevOps teams and the data scientists supporting these teams.

The case of {\bf autonomous system deployments} requires automation as the primary success factor. Even within a single software deployment, autonomous systems can use system logs to retrain the machine learning models frequently. In several cases, the system is able to determine the success of a classification or prediction after some time, which allows it to determine the performance of the models during operation. In general, autonomous systems can detect changes in the parameter values of log entries and identify when the system moves outside of normal behavior. This information can then be used to trigger retraining or to generate a warning that human intervention is required.

In table~\ref{tab:sucfac}, we summarize the scenarios and success factors. It is important to realize that in the case of autonomous systems, humans are out of the operational loop, meaning that effort in that case refers to the effort required to build the automated infrastructure for self-monitoring, automatic retraining and warning generation.

\begin{table}[]
\begin{tabular}{lllll}
                   & Effort & Data Relevance & Automation &  \\
Traditional Logs   & ++     & o              & o          &  \\
DevOps             & +      & ++             & +          &  \\
Autonomous Systems & +      & +              & ++         & 
\end{tabular}
\caption{Scenarios and success factors}
\label{tab:sucfac}
\end{table}

\subsection{Technical Framework}

In this paper, we present an approach for generating system logs that are optimized for machine learning, rather than for interpretation by humans. This approach consists of a technical part and a process part. In this section we present the technical part.

As machine learning algorithms benefit from structure, the log entries making up the system log should preferably follow a predefined syntax. In the approach presented in this paper and based on established best practice in industry, a log entry is organized into two distinct parts, i.e. an {\it identifying } section and a {\it parameter} section. The identifying section can be viewed as a header containing a number of elements:
\begin{itemize}
    \item {\bf Entry type and class}: The type of entry depends on the system and the R\&D organization (see also the next section), but for a typical system, types include info, debug, warning and error. For each type, there are one or more classes that can be reported upon. For instance, for the type warning, there might be classes related to connectivity, storage, out-of-bound data, etc.
    \item {\bf Server and process}: The log entry needs to know the source of the information which typically include the server and process information. Again, the exact data depends on the R\&D organization as well as the deployment at the customer, but the log entry needs to be identifiable from the perspective of the server and process.
    \item {\bf File and location in file}: As the software in most systems is organized in files, relevant identifying information includes the file name and the location in the file. 
    \item {\bf Component and function}: Next to the file structure view, it is also relevant to provide a clear location in the logical view, i.e. the subsystem or component as well as the function or method from which the log entry is generated.
\end{itemize}

As is obvious from the structure of the {\it identifying} part, depending on the use case, machine learning models need to be able to use various identifying data elements as features in order to provide relevant conclusions or inference. For instance, the algorithm may learn that different servers behave differently and that what might constitute an anomaly for one server is actually normal behavior for another one.

\begin{figure}[!ht]
    \centering
    \includegraphics[width=9cm, keepaspectratio]{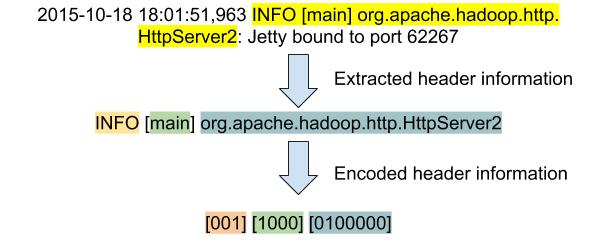}
    \caption{Example numerical/encoded representation of the identifying information of a HDFS (Hadoop File System) log entry}
    \label{fig:header_info}
\end{figure}

To illustrate the identifying part of a log entry, we show an example in figure~\ref{fig:header_info}. In this case, we use a log entry generated by the Hadoop File System (HDFS). In this case, for the header information, we would identify the parts that are identifying, in this case the type of log entry (INFO), the program (main) and the server ID (2). Each of these is encoded using one-hot encoding, resulting in a series of bits representing each of the parts. Rather than generating the text shown at the top of the figure, a log entry optimized for machine learning would be generated along the lines of the bottom row in the figure.

The parameter part consists of zero, one or more data points. In our framework, all data points are numerical, but can either be  continuous or discrete - serving as a unique identifier for some state (e.g. 0001, indicating a one-hot encoding of entry type "Error"). 

As the log file is intended to be used for machine learning without any preprocessing, also the parameter data should be optimally structured. The approach for structuring data well depends on the type of data that is stored as a parameter. Using the levels of measurement defined by Stanley Smith Stevens~\cite{wiki_Stevens}, below we describe for each type of data how the parameter should be provided as part of the log entry.

The first data type is the nominal scale where the different elements on the scale have no order or relationship to each other. The typical approach is to use one-hot encoding~\cite{wiki-OneHot}. One-hot encoding maps a set of alpha-numerical variables to a bit string of the same length as the number of different strings. In that way, each string is mapped to a unique binary number with a single ``1'' bit.

The ordinal scale is the second type of data. In this case, the elements on the scale have a rank order, e.g. 1st, 2nd, 3rd, but the distance or ratio between elements with higher or lower rank is not defined. The ordinal scale can be used for dichotomous data, e.g. ``healthy'' or ``sick'', and for non-dichotomous data, e.g. ``strongly disagree'', ``disagree'', etc. In the case of dichotomous data, the element can be encoded with a single bit. For non-dichotomous data, several alternative types of encoding exist to ensure that machine learning algorithms to use the fact that the scale has a rank order. Examples include Halmert effect encoding, orthogonal polyonomial encoding and difference contrast encoding, but a detailed discussion of these encoding approaches goes beyond the goal of this paper. Instead we refer to~\cite{spss} for a more in-depth discussion.

The other data types, i.e. interval and ratio, both require normalization. Normalization maps numerical values on a random range to values between 0 and 1. The simple approach is to define the minimum and maximum and to calculate the normalized value as {\it normalized value = (original value - minimum value)/maximum value}.

There are two factors that may complicate this approach. First, it is not necessarily known in all cases what the minimum and maximum values are. In this case, the software responsible for generating the associated log entry needs to keep track of the actual value to be normalized and, at run-time, adjust the minimum or maximum value. The challenge is that when this happens, all earlier log entries of this type are no longer valid as the normalized parameters are now off the new scale.

There are three main strategies to address this. First, one can perform a traversal of the system log and re-normalize the values of the affected parameter. Second, one can decide to simply ignore the adjustment under the assumption that the machine learning algorithm will be sufficiently robust to deal with a shift in normalization. Especially in the case when a time window over the data is used, the shift in normalization will be temporary until the last log entry with the old normalization values falls outside the window.

Third, using a more robust and/or specific normalisation technique can be employed. There are several normalisation techniques that do not require hard min-max boundaries to operate correctly. For example, the Sigmoid or tanh functions can map values from $-\infty$ to $\infty$ to a value between 0 (or -1) and 1. However, the success of these techniques is heavily dependent on the parameters and assumptions made. If the original data is of an exponential distribution and a Sigmoid or tanh function is used for normalisation, then much of the variance will be lost. And, similar to the problem in min-max normalisation, consistent changes in parameter values will render the normalisation inaccurate, meaning that adjustment may need to be made.

The second factor is concerned with statistical distribution. While neural networks perform better with standardised input and tend to perform well if the data follows Gaussian distributions, it may also perform well if the data follows other distributions. For example, in our experience, in many cases a binomial distribution performs even better. Other machine learning algorithms like decision trees don't require normalisation, rather it doesn't influence their performance. That said, there are machine learning algorithms that are dependent on the data following a Gaussian distribution. As we we want to avoid a situation where the generated data is only useful for some machine learning algorithms, it is generally advisable to generate system logs such that it is as widely applicable as possible. This may require mapping the actual distribution to the data to a Gaussian one. Doing so minimizes dependencies between data generation and the machine learning models.

\begin{figure}[!ht]
    \centering
    \includegraphics[width=9cm, keepaspectratio]{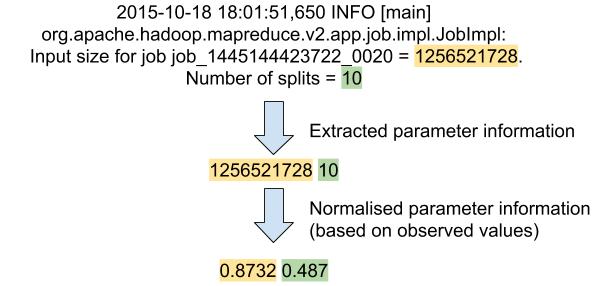}
    \caption{Sample numerical/normalised representation of the parameter information of a HDFS (Hadoop File System) log entry}
    \label{fig:parameter_info}
\end{figure}

For the parameter part of the log entry, we show an example in figure~\ref{fig:parameter_info}. Similar to the previous example, we take a log entry from the Hadoop File System and focus on the parameter part. In case the size for an input job (1256521728) and the number of splits (10). Both of these values need to be normalized and for this we can analyze the typical values for input job and number of splits. In this case, the (sample) normalization of the input job size turns out to be 0.8732 and for the number of splits to be 0.487.

\subsection{Process Approach}

In this paper we discuss the generation of system logs that are optimized for machine learning. This approach contains a technical framework that we discussed in the previous section. This needs to be complemented with a process approach to ensure that the R\&D organization operates in accordance to the preconditions for the technical approach.

It is important to note that the process approach that we describe in this section is only concerned with the generation of system logs for machine learning. It does not prohibit the organization to continue to use other types of system logs that are used for other purposes. However, if the intent is to use a certain system log for machine learning, the technical framework outlined in the previous section facilitates the use of logs that require no pre-processing. To achieve this, there are a number of process activities that need to be governed across the R\&D organization in order to ensure generation of suitable log entries. These processes include five main activities.

\subsubsection{Definition of log entry creation policy}

The first activity is to define a policy for the generation of log entries. In practice, we see that in many companies, there are lots of different levels of granularity for log entries. Some generate frequently and periodically whereas others only generate entries when infrequent events happen to the system. Our general guidance is that log entries should be generated for relevant system states and significant system events, but the actual interpretation for a specific system should be defined by the R\&D organization for the system. The main concern is, however, that all teams use the same policy so that log entries generated by different teams can still be used in the same way by machine learning algorithms.
    
\subsubsection{Governance of identifying variables and encoding strategy}

For each of the variables in the identifying part of the log entry, e.g. entry type, component name, etc., the R\&D organization needs to define the specific names that are to be used and the way that these names are encoded (typically using one-hot encoding). 

As the system will be deployed to different customers, the organization needs to decide the encoding of the identifying variables when it concerns servers and processes in order to be able to distinguish between system logs from different customers while still being able to combine logs from these customers for training of machine learning models.

\subsubsection{Selection of parameter normalization and encoding approach}

For the parameters associated with each type of log entry, the normalization and encoding strategy needs to be defined and enforced by the organization. As discussed in the technical framework, parameters typically are on the nominal, ordinal, interval or ratio scale. For each parameter, first the type of data needs to be identified. Based on this, the organization needs to define the encoding strategy. Finally, for parameters of interval and ratio, it may be necessary to consider the statistical distribution of the data and, if necessary, provide a mapping from the existing statistical distribution to a Gaussian one. 

The process of encoding, normalizing and mapping distributions can be quite complex and require experimentation. In the case of a DevOps approach, it may be necessary to deploy the relevant code, generate system logs and analyze these solely for the purpose of verifying that the data contains sufficient information to use for machine learning purposes.

\subsubsection{Selection of labeling strategy}

As the vast majority of machine learning algorithms use supervised learning approaches, it is important to make sure that relevant labels are associated with each log entry. Sometimes, log entries are intrinsically labeled by their structure, e.g. a log entry of type ''error'' and class ''no connection'' has a clear label associated. In other cases, labels can be associated with a log entry after a period of time, e.g. when a machine learning model makes a prediction, after some time it often is possible to record the actual outcome and to associate this outcome with the log entry as a label. Finally, it may be necessary to use manual labeling. This is acceptable for the traditional and, to a lesser extent, DevOps scenarios, but not for autonomous system deployments.

\subsubsection{Governance of evolution and backward compatibility}

Finally, once the initial structure of log entry types and parameters is in place, there will be a constant flow of change requests from the R\&D teams, customers, data scientists and others. The evolution of the system log entry model needs to be carefully managed as allowing for breaking changes may also invalidate data sets predating the change due to changes in semantics and/or structure.

In the cases where introducing breaking changes is unavoidable, it may be necessary to develop mapping functions that allow for the generation of data sets that are based on system logs both before and after the breaking change. As virtually all machine learning algorithms perform better with a greater quantity of data, it is frequently beneficial to combine multiple logs into one data set for training and validation.

\subsection{Summarizing}

\begin{figure}\centering
\includegraphics[width=9cm, keepaspectratio]{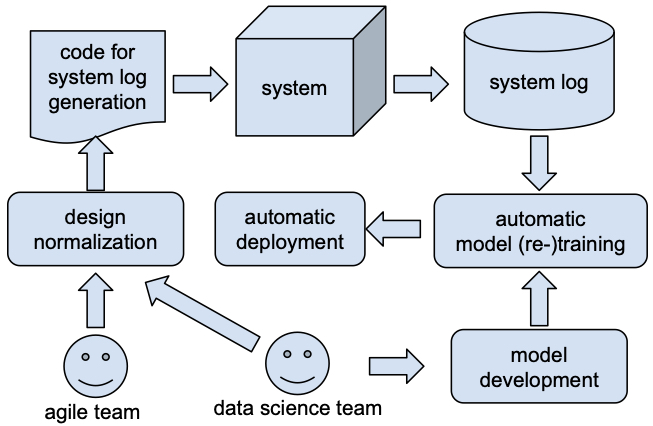}
\caption{System logging for machine learning}
\label{MLSysLog}
\end{figure}

As presented in this section and in figure~\ref{MLSysLog}, generating logs for machine learning requires that engineers, R\&D teams and the organization change the way log entries are generated. However, the process by which system logs for machine learning are generated is, in principle, no more difficult than adding a normal log statement. The main difference is the organizational alignment and agreement on the structure and semantics of log entries. As usual, although most of the attention is quickly drawn towards the technical framework, it is the introduction of new processes and activities that will require the most effort and attention.

Especially early in the process of adopting the approach outlined in this paper, it is beneficial to add a new AI log statement in the code at every place that there is an existing log statement, leaving the existing log statement. In this way, it is possible to generate two separate logs: the original log and the new AI log. The information that is presented in the AI log statement should be an encoded/normalised version of what is presented in the human readable log. In this manner, it will be very simple to link a human-readable log entry with an AI log entry. In the case that a machine learning algorithm finds an anomaly in the AI log, the link with the human understandable entry in the human-readable log significantly helps the investigation into the detected anomaly.

If, when optimising the AI log input, there are cases where log statements should be added or removed, the link between an AI log entry and a human log section (or entry) should, where possible, remain. For both the developer and the data scientist, the results of machine learning should be, at the very least, actionable and, preferably, explainable. Parallel machine and human readable logs provide a very good way of achieving that.


Managing the evolution of logs over time is perhaps the most difficult problem in the process. Both the content of log entries as the order in which log entries are generated influences the training of machine learning models. One obvious method to hinder the effects is to train models in an online manner. However, this typically assumes the autonomous system scenario presented earlier. The challenge here is that the product is deployed at several or many customers. Each local incarnation of the system may do its own run-time training, but combining the learnings of many deployed systems into a generalized model and redeploying this model, at run-time, to all the deployments of the system is a very hard task that most companies have decided to not (yet) take on. Local optimization of machine learning models, however, could benefit from local indications relevant for the model, such as changes in software or hardware configuration. 

\section{Validation}\label{validation}

To validate the approach presented in the paper, we conducted a validation workshop at the case study company during January 2020. A total of eight domain experts participated. The validation workshop consisted of four main parts, concerning the four main contributions of this paper, i.e. the problem statement, the scenarios and success factors, the technical framework and the solution approach. After the workshop, there were follow-up meetings with some of the participants to deep-dive on the technical framework and the process approach with the intent of applying this in one of the products developed by the case study company. The results from the validation with the R\&D teams are not yet available at the time of submitting this paper, so this section focuses on the validation workshop and the follow-up meetings.

\subsection{Problem statement}
In section~\ref{problem} we presented eight challenges associated with the contemporary way in which system logs are generated. When presenting these challenges, we received complete and unreserved support from the domain experts. Each of the domain experts had experienced most or all of the challenges that we identified and recognized these without any prompting.

When asked whether we had missed any challenges, there was discussion among the experts but no new challenges were identified. Our conclusion is that the problem statement accurately represents the challenges associated with system logs at the case study company.

\subsection{Scenarios and success factors}
During the validation workshop, we presented the three scenarios, i.e. traditional log analysis, DevOps context and autonomously improving systems. The domain experts recognized the first two scenarios, especially as the company has shifted from traditional deployments to a DevOps model during recent years.

The notion of autonomously improving systems was discussed and although the domain experts recognized this as a likely future scenario, no instances of this scenario were currently in operation at the case study company. When asked to hypothesize on the requirements on generating system logs for this scenario, the domain experts confirmed that the approach presented in the paper contained the necessary ingredients to support this case.

The success factors were confirmed as relevant, but the domain experts added the notion of computational efficiency as a factor to consider.

\subsection{Technical framework}

The domain experts confirmed the {\it identifying} part of the technical framework without any further comments. The parameter part of the framework, however, led to significant discussion. The main concern was that encoding, normalization and mapping between statistical distributions caused the data to be lost in the process. 

In response to our suggestion that maintaining a parallel system log for human interpretation could be used for that purpose, the domain experts were sceptical as, in practice, they experience many challenges and manual effort associated with these logs. Consequently, the overall consensus among the domain experts was to include the original data in the log entry as well.

The above led to a discussion concerning the goal of the model to avoid all pre-processing of system logs. This caused an interesting observation that the domain experts would like to automate 95\% of the process, but keep the last 5\% for themselves. The overall conclusion by us was that the desirability of avoiding all pre-processing is not universally shared by data scientists as this represents that vast majority of their current job responsibilities.

\subsection{Process approach}

The domain experts confirmed that the process activities that we identified were correct and offered no other activities to be added or existing ones removed. The main discussion was concerned with how to influence the R\&D organization to actually implement these changes. 

The domain expert working closest with the product teams was less concerned than the others who were less connected. However, the general consensus was that significant change management activities were required for the product organizations to adopt the approach, especially as it requires significant alignment and governance across the organization and constrains the freedom of agile teams.

Concluding, the validation workshop and follow-up meetings confirmed our approach. There were some smaller concerns raised concerning the approach, but the change management required to realize the approach was viewed as the main obstacle.

\section{Discussion}\label{discussion}

In section~\ref{problem}, we presented eight challenges associated with contemporary approaches to system log generation. The approach that we present in this paper aims to address these challenges.

For challenges (i), (vi), (vii), and (viii), we proposed that system logs for machine learning should be standardised. Each log entry should have the exact same format and should be completely numerical. To avoid problems with variable length data, every log entry of the same type should have the same number of parameters. In the cases of dynamic parameter information - where an indeterminate number of processes or parameters may spawn - we recommend the developer find a way to reduce the dimensionality to the same number of parameters (e.g. just print the number of processes spawned, rather than each of their names). Despite the debugging purposes the display of this dynamic information may contain, it serves little use for machine learning algorithms, which rarely take variable input quantity.

By generating all data in a numerical and standardised form, we address the first challenge (i). By forcing this standardisation on the developer, we don't allow them to make new formats. As an additional advantage, by doing so we also eliminate the case that log file interpreters and parsers fail unpredictably (viii).

As a solution to challenge (iii), we propose that all information should write into the same log file. The principle behind the separation of domain and level of abstraction is fair, but we argue that big data is greater than the separation of data (as similar results have been shown in earlier studies~\cite{Banko2001}).

Arguably the most difficult challenges to solve are (ii) and (iv), because, by the combination of separate log files, differing levels of abstraction and multiple processes writing into the same log file would be emphasized. It is from this that we propose standardisation across levels of abstraction by matching the lowest common denominator. With all subsystem processes writing into the same file, standardisation (from a complexity and time perspective) would occur naturally. This is rationalised through the same principle of big data being better than separated, but cleaner data~\cite{Banko2001}.

\subsection{Threats to validity}

In empirical software engineering research, we typically recognize four types of threats to validity: conclusion, internal, construct and external~\cite{feldt10}. 

The validity of the conclusion of our research is warranted by the validation workshop with several domain experts at the case study company as well as follow-up meetings. However, the approach has not yet been tested in the context of a deployed product.

We have sought to warrant against threats to the internal and construct validity of the research by studying system logs for two generations of systems, two different subsystems, two types of logs (error and operational) and two different contexts (internal testing and at the customer). 

We do recognize that there are threats to external validity as the research is based on a single case study company. Although our experience from other companies is largely in line with the findings with our case study company, we do not present those findings in this paper.

Although we recognize that some threats to validity exist, we are confident that the approach presented in this paper is applicable to many more systems across a variety of industries as the problems that we identified are in no way specific for the case study company and the domain in which it operates.


\section{Conclusion \& Future work}\label{conclusion}

System logs perform a critical activity in software-intensive systems as these record the state of the system and significant events in the system at important points in time. This facilitates significantly simplified defect management, anomaly detection, monitoring of system performance over time and even prediction of future system behavior. As a consequence, the use of system logs is ubiquitous for software intensive systems and the exchange of log files between users of these systems and the organizations that build them is a typical activity.

Despite their ubiquitous and highly informative nature, the full potential of system logs is not realized because although most logs are intended for human interpretation, the constantly increasing system size and complexity has caused these logs to have an unmanageable size. This calls for the use of machine learning techniques, but the unstructured, alphanumerical and human oriented nature of logs limits the applicability of machine learning algorithms as there typically is high levels of noise and irrelevant information in system logs. This causes a need for significant pre-processing of logs which requires large amounts of human effort, reducing the benefits that fully automated machine learning could provide.

In this paper, we presented a technical framework and process approach for generating system logs that addresses the aforementioned challenges, removes the need for pre-processing and opens up for the direct use of system logs for training of machine learning models and inference based on the information in logs.

The contribution of this paper is threefold. First, we presented the main challenges of contemporary approaches to generating and storing log data for large, complex, software-intensive systems based on an in-depth case study at a world-leading telecommunications company. Second, we presented a systematic and structured approach for generating log data that does not suffer from the aforementioned challenges and that allows for immediate use by machine learning algorithms. Third, we provided validation based on expert interviews that confirms that the approach addresses the identified challenges and problems.

As part of our future work and as discussed in the paper, we plan to apply the logging approach with a group of agile development teams at the case study company. In addition, we aim to reach out to other companies to confirm our findings and approach.

\section*{Acknowledgment}
This research has in part been funded by Vinnova in the context of the ``HoliDev - Holistic DevOps Framework'' project.

\bibliographystyle{IEEEtran}

\end{document}